\documentclass[aps,floats]{revtex4}
\usepackage{amsmath,amssymb}
\usepackage{graphicx,epsfig}

\begin{document}
\bibliographystyle {plain}

\def\oppropto{\mathop{\propto}} 
\def\opsimeq{\mathop{\simeq}}
\def\opoverderline{\mathop{\overline}}
\def\operarrow{\mathop{\longrightarrow}}
\def\opsim{\mathop{\sim}}

\def\fig#1#2{\includegraphics[height=#1]{#2}}
\def\figx#1#2{\includegraphics[width=#1]{#2}}


\title{ Many-Body-Localization Transition : \\
 sensitivity to twisted boundary conditions  } 


 \author{ C\'ecile Monthus }
  \affiliation{ Institut de Physique Th\'{e}orique, 
Universit\'e Paris Saclay, CNRS, CEA,
 91191 Gif-sur-Yvette, France}

\begin{abstract}

For disordered interacting quantum systems, the sensitivity of the spectrum to twisted boundary conditions depending on an infinitesimal angle $\phi$ can be used to analyze the Many-Body-Localization Transition. The sensitivity of the energy levels $E_n(\phi)$ is measured by the level curvature $K_n=E_n''(0)$, or more precisely by the Thouless dimensionless curvature $k_n=K_n/\Delta_n$, where $\Delta_n$ is the level spacing that decays exponentially with the size $L$ of the system. For instance $\Delta_n \propto 2^{-L}$ in the middle of the spectrum of quantum spin chains of $L$ spins, while the Drude weight $D_n=L K_n$ studied recently by M. Filippone, P.W. Brouwer, J. Eisert and F. von Oppen [arxiv:1606.07291v1] involves a different rescaling. The sensitivity of the eigenstates $\vert \psi_n(\phi) > $ is characterized by the susceptibility $\chi_n=-F_n''(0)$ of the fidelity $F_n =\vert < \psi_n(0) \vert \psi_n(\phi) >\vert $. Both observables are distributed with probability distributions displaying power-law tails $P_{\beta}(k) \simeq A_{\beta} \vert k \vert^{-(2+\beta)} $ and $Q(\chi)  \simeq B_{\beta} \chi^{-\frac{3+\beta}{2}} $, where $\beta$ is the level repulsion index taking the values $\beta^{GOE}=1$ in the ergodic phase and $\beta^{loc}=0$ in the localized phase. The amplitudes $A_{\beta}$ and $B_{\beta}$ of these two heavy tails are given by some moments of the off-diagonal matrix element of the local current operator between two nearby energy levels, whose probability distribution has been proposed as a criterion for the Many-Body-Localization transition by M. Serbyn, Z. Papic and D.A. Abanin [Phys. Rev. X 5, 041047 (2015)].

\end{abstract}

\maketitle

\section{ Introduction }

The sensitivity to boundary conditions has been used in many areas of physics to  distinguish different phases of matter.
The idea is that the presence of a finite correlation length $\xi$ implies that the effect of boundary conditions decays exponentially
with respect to the system size $L$.
This criterion turns out to be especially interesting in the field of disordered systems whenever the local order is not
so easy to define and varies from sample to sample.
For instance for classical spin-glasses, the free-energy difference between periodic and anti-periodic boundary conditions
in a given disordered sample allows to characterize the low-temperature spin-glass phase by its droplet exponent 
\cite{mcmillan,bray_moore,fisher_huse} and to distinguish it from the high-temperature paramagnetic phase.

 For quantum systems, one may also consider the difference between periodic and antiperiodic boundary conditions \cite{edwards_thouless}, 
but it can be more convenient to consider twisted boundary conditions
depending on an infinitesimal angle $\phi$ in order to be able to follow continuously the spectrum \cite{kohn}. 
Then the sensitivity of each energy level 
\begin{eqnarray}
E_n(\phi)  = E_n(0)+ \frac{K_n}{2} \phi^2 + o(\phi^2)
\label{Kbig}
\end{eqnarray}
is measured by the level curvature $K_n$. Its statistical properties have been much studied in
chaotic quantum systems and random matrices \cite{gaspard,delande,vonOppen,fyodorov_sommers_rmt,fyodorov_edge} and in Anderson Localization models
\cite{mcmillan_anderson,fyodorov_sommers,kravtsov,fyodorov,montambaux,kramer,evangelou}. The case of the ground state energy in 
interacting quantum models has been analyzed both without disorder \cite{fye,henkel}
and with disorder \cite{bouzerar,rieger,saha}, in particular to characterize the persistent currents induced by a magnetic flux in mesoscopic rings.
 The case of highly excited eigenstates 
in a disordered model of interactions fermions has been investigated very recently \cite{eisert}
in the context of the Many-Body-Localization problem
(see the recent reviews \cite{revue_huse,revue_altman} and references therein).
In the field of Many-Body-Localization, the sensitivity of energy levels to a local perturbation of finite amplitude
has been already proposed as a criterion for the MBL transition in \cite{serbyn_criterion}
and discussed from the point of view of finite-size scaling in \cite{c_mblpseudocriti}.
The closely related problem of the coupling between two long chains has been also discussed in \cite{serbyn_criterion,c_mblrgeigen}.
One goal of the present paper is thus to make the link between these works concerning 
a local perturbation of finite amplitude \cite{serbyn_criterion,c_mblpseudocriti,c_mblrgeigen},
and the infinitesimal perturbation introduced by the twist angle $\phi$ (Eq. \ref{Kbig}) considered in \cite{eisert}.

Besides the sensitivity of energy levels, it is also interesting to consider 
the sensitivity of the corresponding eigenstates.
 The fidelity,
defined as the overlap between the unperturbed eigenstate $\vert \psi_n(\phi=0) > $ and the perturbed eigenstate $ \vert \psi_n(\phi) >$ 
(see the review \cite{review_fidel} and references therein)
can be also expanded with respect to the twist angle $\phi$ as
\begin{eqnarray}
F_n(\phi)  \equiv \vert < \psi_n(0) \vert \psi_n(\phi) >\vert =1- \frac{\chi_n}{2} \phi^2 + o(\phi^2)
\label{chifidel}
\end{eqnarray}
The fidelity susceptibility $\chi_n $ has been much considered in the context of quantum phase transitions
without disorder \cite{zanardi,saleur,dubail,konig}. This question of sensitivity of eigenstates is of course closely related to the 
famous Anderson orthogonality catastrophe \cite{anderson_orthog}, 
which has been also studied in various disordered models,
in particular in random matrices \cite{aleiner,gefen},
in Anderson Localization models \cite{altschuler,kettemann},
at random critical points at zero temperature \cite{vasseur_orthog},
and in Many-Body-Localization models \cite{sondhi,sarma}.

From a physical point of view, one expects that the sensitivity of energy levels (Eq. \ref{Kbig})
and of eigenstates (Eq. \ref{chifidel}) should be in direct correspondence, so that in the present paper we wish to analyze both in parallel
for Many-Body-Localization models. Our main conclusion is that both probability distributions display power-law tails with exponents involving the level repulsion index $\beta$ taking the values $\beta^{GOE}=1$ in the ergodic phase and $\beta^{loc}=0$ in the localized phase, while the amplitudes of these two heavy tails are given by moments of the off-diagonal matrix element of the local current operator between two nearby energy levels, whose probability distribution has been proposed as a criterion for the Many-Body-Localization transition \cite{serbyn_criterion}.

The paper is organized as follows.
In section \ref{sec_twist}, the effect of twisted boundary conditions 
is described on the example of the XXZ quantum spin chain with random fields,
 in order to obtain the level curvatures of the eigenvalues and the fidelity susceptibilities of the eigenstates.
Section \ref{sec_curvature} is devoted to the probability distribution of
the Thouless dimensionless level curvature,
while section \ref{sec_chi} focuses on the probability distribution of the fidelity susceptibility $\chi_n$
of eigenstates. In section \ref{sec_ampli}, the scaling with the system size $L$
of the amplitudes of the heavy tails of the two probability distributions is analyzed.
Section \ref{sec_conclusion} summarizes our conclusions.

\section{ Many-Body-Localization models with twisted boundary conditions }

\label{sec_twist}

\subsection{ Random field quantum spin chain with twisted boundary conditions }

The XXZ quantum spin chain with random fields $h_j$ 
\begin{eqnarray}
H_0= J \sum_{j=1}^{L} \left(\sigma_j^{+} \sigma_{j+1}^- + \sigma_{j}^-\sigma_{j+1}^{+}  \right)
+  \sum_{j=1}^L \left( h_j \sigma_j^z + J^{zz} \sigma_j^z \sigma_{j+1}^z \right)
\label{h0}
\end{eqnarray}
is the most studied Many-Body-Localization model where numerical results for many observables are available
\cite{kjall,alet,alet_dyn,luitz_tail,badarson_signa,auerbach,znidaric_dephasing,prelo_dyn,znidaric_lindblad,serbyn_powerlawent,luitz_bimodal,garcia}

As explained in the Introduction, 
besides the usual periodic boundary conditions $\sigma_{L+1}=\sigma_{1} $ , it is interesting to
consider twisted boundary conditions with some angle $\phi$ for the spin operators \cite{rieger}
\begin{eqnarray}
\sigma_{L+1}^{\pm} && = e^{ \pm i \phi} \sigma_1^{\pm} 
\nonumber \\
\sigma_{L+1}^{z} && =  \sigma_1^{z} 
\label{bcflux}
\end{eqnarray}
In terms of the Hamiltonian, 
this means that Eq. \ref{h0} is changed into
\begin{eqnarray}
H(\phi) && = H_0 + V(\phi)
\label{hphi}
\end{eqnarray}
where the new contribution involves only the two boundary spins $ \sigma_1$ and $\sigma_L $ \cite{rieger}
\begin{eqnarray}
V(\phi) && \equiv J (e^{i \phi}-1) \sigma_{1}^{+}  \sigma_{L}^{-} +
J (e^{-i \phi}-1)\sigma_{1}^{-} \sigma_{L}^{+}
\label{vphi}
\end{eqnarray}

Here we should stress the difference with the recent study \cite{eisert}
where the same model is written in the language of interacting fermions (Jordan-Wigner transformation),
so that the twist angle $\phi$ has the physical meaning of a magnetic flux : 
the authors of  \cite{eisert} have chosen to work in the gauge where 
the whole flux $\phi$ is taken into account by affecting $\frac{\phi}{L}$ to each of the $L$ bonds,
as also done in \cite{bouzerar,saha}, while here we have chosen to work in the gauge 
where the whole flux $\phi$ is affected to the single bond between $ \sigma_1$ and $\sigma_L $ as in \cite{rieger}.
Although the two choices are equivalent via a gauge transformation, 
the present choice where the perturbation of Eq. \ref{vphi} is local and involves a single bond
is much easier to analyze than the case where the perturbation is spread over all bonds,
and allows to make a direct link with previous works based 
on the effects of  local perturbations in Many-Body-Localization models \cite{serbyn_criterion,c_mblpseudocriti,c_mblrgeigen}.

\subsection{ Second Order Perturbation Theory in the twist angle $\phi$ }

The expansion up to second order in $\phi$ of the contribution of Eq. \ref{vphi} to the Hamiltonian (Eq. \ref{hphi})
\begin{eqnarray}
V(\phi) && =  - \phi {\cal I } - \frac{\phi^2}{2} {\cal T}  +o(\phi^2)
\label{hamiltonphiexpansion}
\end{eqnarray}
involves at first order the local boundary current operator ${\cal I} $ 
\begin{eqnarray}
{ \cal I}  && \equiv  - i  J (\sigma_{1}^{+}  \sigma_{L}^{-}- \sigma_{1}^{-}\sigma_{L}^{+}  ) 
\label{Ioperatot}
\end{eqnarray}
and at second order the local boundary kinetic energy operator ${\cal T} $ 
\begin{eqnarray}
{\cal T} && \equiv J ( \sigma_{1}^{+}  \sigma_{L}^{-}+\sigma_{1}^{-} \sigma_{L}^{+} )
\label{toperator}
\end{eqnarray}

Let us introduce the spectral decomposition of the unperturbed Hamiltonian $H_0$ (Eq. \ref{h0})
corresponding to periodic boundary conditions $\phi=0$
\begin{eqnarray}
 H_0  = \sum_{n=1}^{2^L} E_n(0) \ \vert \psi_n(0)  >< \psi_n(0) \vert
\label{h0spectral}
\end{eqnarray}
in terms of the $2^L$ eigenvalues $E_n(0)$ and the associated eigenvectors $\vert \psi_n(0) >$.

The second order perturbation theory in the twist angle $\phi$ for the eigenvalues 
(the first order correction vanishes $< \psi_n(0) \vert {\cal I}   \vert \psi_n(0) >=0 $ as a consequence of the time-reversal symmetry of $H_0$)
yields that the level curvature of Eq. \ref{Kbig} reads
\begin{eqnarray}
K_n && = - < \psi_n(0) \vert {\cal T}   \vert \psi_n(0) >
+2 \sum_{m \ne n} 
\frac{ \vert < \psi_m(0) \vert {\cal I}  \vert \psi_n (0)> \vert^2 } {E_n(0)-E_m(0)}
\label{Knper}
\end{eqnarray}
The first contribution involves the local boundary kinetic energy of the unperturbed eigenstate $< \psi_n(0) \vert {\cal T}   \vert \psi_n(0) > $, while the second contribution involves the energy differences $(E_n(0)-E_m(0))$ and the off-diagonal matrix elements $< \psi_m(0) \vert {\cal I}  \vert \psi_n (0)> $ of the local boundary current operator.

The second order perturbation theory for the eigenvectors yields
that the fidelity susceptibility of Eq. \ref{chifidel}
\begin{eqnarray}
\chi_n =  \sum_{m \ne n}  \left\vert\frac{  < \psi_m(0) \vert {\cal I}  \vert \psi_n (0)> } {E_n(0)-E_m(0)} \right\vert^2
\label{chinper}
\end{eqnarray}
also involves energy differences and the off-diagonal matrix elements of the local boundary current operator ${\cal I} $.

The comparison between Eqs \ref{Knper} and \ref{chinper} shows the close relationship between the two,
although the fidelity susceptibility $\chi_n$ is somewhat simpler for two reasons :
 it is dimensionless and it corresponds to a sum of positive contributions, while the 
curvature $K_n$ has the dimension of an energy 
and corresponds to a sum of positive and negative contributions.
Nevertheless, since the level curvatures have been much more studied in disordered systems,
we will focus first on the probability distribution of the level curvature in the next section.

\section{ Statistical properties of
 the dimensionless curvature $k_n$ of energy levels }

\label{sec_curvature}

\subsection{ Thouless dimensionless curvature $k_n$ involving the level spacing }

Since the curvature $K_n$ has the dimension of an energy by definition  (Eq. \ref{Kbig}),
it is essential to introduce the Thouless dimensionless curvature 
\begin{eqnarray}
k_n \equiv \frac{K_n}{\Delta_n} 
\label{ksmall}
\end{eqnarray}
where $\Delta_n$ is the typical level spacing around the energy level $E_n$.
This rescaling by $ \Delta_n$ has been first introduced by Edwards and Thouless \cite{edwards_thouless}
in the context of the Anderson Localization where it is used systematically \cite{mcmillan_anderson,fyodorov_sommers,kravtsov,fyodorov,montambaux,kramer},
and where it has played a very essential role in the elaboration of the scaling theory of localization \cite{abrahams}.

So here we should stress an important difference between Anderson Localization models
and Many-Body-Localized models :

(i) in Anderson Localization (AL) models for a single particle, a system of $L^d$ sites 
contains $L^d$ eigenstates, so that the level spacing behaves as the power-law
\begin{eqnarray}
\Delta_n^{(AL)} \propto L^{-d}
\label{deltaAnderson}
\end{eqnarray}

(ii) in Many-Body-Localization models of $L$ spins containing $2^L$ eigenstates as Eq. \ref{h0},
the level spacing at a given energy-density $e=\frac{E_n}{L}$ per spin
decays exponentially with the size $L$
\begin{eqnarray}
\Delta^{(MBL)}_n=\Delta(E_n=Le)  \propto   e^{-L S(e) }
\label{largedev}
\end{eqnarray}
with some entropy function $S(e)$. In particular near the middle of the spectrum $e=0$,
the level-spacing follows the Gaussian small-deviation regime
\begin{eqnarray}
\Delta^{(MBL)}_n=\Delta(E_n=Le)  \simeq \ \sqrt { 2  \pi L \sigma^2 }  \   e^{-L \left( \ln 2 -  \frac{ e^2}{2  \sigma^2}  \right) }
\label{gaussdelta}
\end{eqnarray}
in all nearest-neighbor quantum spin chains \cite{atas,keating} and in particular for the MBL  spin chain of Eq. \ref{h0} \cite{c_mblpseudocriti}.

So the dimensionless curvature $k_n$ of Eq. \ref{ksmall}
based on the rescaling with the typical level spacing $\Delta_n$ involves a very different rescaling with the size $L$
from the 
Drude weight $D_n$ studied in \cite{eisert}
\begin{eqnarray}
D_n = L K_n = L \Delta_n k_n \simeq \sqrt { 2  \pi \sigma^2 }  \   L^{\frac{3}{2} }  \  e^{-L \left( \ln 2 -  \frac{ e^2}{2  \sigma^2}  \right) } k_n
\label{drude}
\end{eqnarray}

\subsection{ Heavy-tails of the distribution of the dimensionless curvature $k_n$ }

Since the level curvature of Eq. \ref{Knper} involve all eigenvalues $E_m(0)$
 and all matrix elements $< \psi_m(0) \vert {\cal I}  \vert \psi_n (0)> $,
its statistics for an interacting disordered model like the quantum spin chain of Eq. \ref{h0}
cannot be computed explicitly. However, it has been understood
on the examples of Anderson localization models \cite{edwards_thouless,mcmillan_anderson,kravtsov,montambaux,kramer,evangelou}
and of quantum chaotic systems and random matrices \cite{gaspard,delande,vonOppen,fyodorov_sommers_rmt}
that the probability distribution of the dimensionless curvature $k_n$
contains some singular power-law tail involving the level repulsion exponent $\beta$ 
\begin{eqnarray}
P^{sing}_{\beta}(k)  \opsimeq_{\vert k \vert \to +\infty} \frac{A_{\beta} }{\vert k \vert^{2+\beta}}
\label{pktail}
\end{eqnarray}
with some amplitude $A_{\beta}$,
so that all non-integer moments $\int dk \vert k \vert^q P(k) $ of index $q \geq 1+\beta$ do not exist.

The presence of this heavy tail can be explained as follows :
a very large value $\vert k_n \vert \to +\infty$ of the dimensionless curvature $k_n$
can only be produced by a very small eigenvalue spacing
$\vert E_{n}-E_{next} \vert =s \Delta_n  $ with $s \to 0$ in Eq. \ref{Knper},
i.e. it occurs near avoided level crossings.
Then the corresponding term dominates over the others in Eq. \ref{Knper},
and the dimensionless curvature can be approximated by this biggest term
\begin{eqnarray}
k_n && =\frac{K_n}{\Delta_n}  \opsimeq_{ s \to 0} 
\frac{ \vert < \psi_{next}(0) \vert {\cal I}  \vert \psi_n(0) > \vert^2 }
{ \Delta_n^2 s  }
\label{knssmall}
\end{eqnarray}
From the small-$s$ behavior of the probability distribution $R_{\beta}(s)$
of the rescaled level-spacing $s$  that involves the level repulsion exponent $\beta$
\begin{eqnarray}
R_{\beta}(s) \oppropto_{s \to 0} s^{\beta}
\label{psbeta}
\end{eqnarray}
and from the probability distribution $\Pi_{n}(w_n)$ of the variable
\begin{eqnarray}
w_n && \equiv  \frac{ \vert < \psi_{next}(0) \vert {\cal I}  \vert \psi_n(0) > \vert^2 }{\Delta_n^2}
\label{wn}
\end{eqnarray}
one obtains by the change of variables of Eq. \ref{knssmall}
\begin{eqnarray}
P^{sing}_{\beta}(k)  && \opsimeq_{\vert k \vert \to +\infty} \int_0^{+\infty} dw \Pi_{n}(w) \int_0^{+\infty} ds R_{\beta}(s)
\delta\left( \vert k \vert -\frac{ w } {  s  } \right)
\nonumber \\
&& \opsimeq_{\vert k \vert \to +\infty} \int_0^{+\infty} dw \Pi_{n}(w) \int_0^{+\infty} ds R_{\beta}(s)
\frac{ \delta\left( s-\frac{ w } { \vert k \vert  } \right) }{ \frac{w}{ s^2} }
\nonumber \\
&& \opsimeq_{\vert k \vert \to +\infty} \int_0^{+\infty} dw \Pi_{n}(w)\frac{ w}{   k^2 }
  R_{\beta} \left(\frac{ w } {  \vert k \vert  } \right)
\nonumber \\
&& \opsimeq_{\vert k \vert \to +\infty} \int_0^{+\infty} dw \Pi_{n}(w)\frac{ w}{  k^2 }
   \left(\frac{ w } {  \vert k \vert } \right)^{\beta}
\nonumber \\
&& \opsimeq_{\vert k \vert \to +\infty} \frac{A_{\beta} }{\vert k \vert^{2+\beta}}
\label{pktailcal}
\end{eqnarray}
This corresponds to the asymptotic power-law behavior announced in Eq. \ref{pktail}
with the amplitude
\begin{eqnarray}
A_{\beta}=  \int_0^{+\infty} dw \Pi_{n}(w) w^{1+\beta}
\label{amplibeta}
\end{eqnarray}
given by the moment of order $(1+\beta)$ of the random variable of Eq. \ref{wn}.

At this stage, it should be stressed that this analysis based only on the statistics on the biggest term in the sum of Eq. \ref{Knper}
allows to predict the asymptotic power-law of Eq. \ref{pktail} for $\vert k \vert \to +\infty $,
but a priori does not give any information on the shape of the distribution for finite values of $k$.
So in general, it will be convenient to decompose the full distribution $P^{full}(k)  $
 into some unknown regular part $P^{reg}(k) $ 
describing the central part of the distribution 
and into the singular power-law tail $P^{sing}_{\beta}(k) $ of Eq. \ref{pktail}
\begin{eqnarray}
P^{full}(k) = P^{reg}(k)+P^{sing}_{\beta}(k)
\label{pkfull}
\end{eqnarray}

\subsection{ Many-Body-Delocalized ergodic phase $\beta=1$ } 

For Many-Body-Localization models, 
the delocalized ergodic phase is governed by the GOE level statistics with the linear level repulsion exponent $\beta^{GOE}=1$,
so that the singular power-law tail of Eq. \ref{pktail} reads
\begin{eqnarray}
P^{sing}_{ergodic(\beta=1)}(k) dk \opsimeq_{\vert k \vert \to +\infty} \frac{A^{ergo}_1 dk }{\vert k \vert^{3}}
\label{pktailgoe}
\end{eqnarray}
where the amplitude (Eq. \ref{amplibeta}) is given by the second moment
\begin{eqnarray}
A_{1}^{ergo}= \int_0^{+\infty} dw \Pi_{n}^{ergo}(w) w^2 
\label{amplibeta1}
\end{eqnarray}

For Random Matrices Ensembles corresponding to the values $\beta=1,2,4$,
 the power-law exponent of Eq. \ref{pktail} has been confirmed by the exact
results for the {\it full } distribution of the appropriate rescaled variable $\kappa$ \cite{vonOppen,fyodorov_sommers_rmt} 
\begin{eqnarray}
P^{RMT}_{\beta}(\kappa) = C_{\beta} \frac{ 1  }{ (1+\kappa^2)^{\frac{2+\beta}{2}}}
\label{rmtbeta}
\end{eqnarray}
with some numerical normalization constant $C_{\beta}$.
(Besides this results for bulk eigenvalues, there also exists results for edges eigenvalues \cite{fyodorov_edge}.)
In Ref. \cite{eisert}, the authors have found numerically that the ergodic phase of the Many-Body-Localization model
is "in excellent agreement" non only with the tail of Eq. \ref{pktailgoe}, but also with the RMT result of Eq. \ref{rmtbeta} with $\beta=1$ for the full distribution.

\subsection{ Many-Body-Localized phase $\beta=0$  } 

The Many-Body-Localized phase is characterized by the Poisson level statistics with no level repulsion $\beta=0$,
so that  the singular power-law tail of Eq. \ref{pktail} reads
\begin{eqnarray}
P^{sing}_{MBL(\beta=0)}(k)  \opsimeq_{\vert k \vert \to +\infty} \frac{A^{MBL}_{0} }{k^{2}}
\label{pktailpoisson}
\end{eqnarray}
where the amplitude (Eq. \ref{amplibeta}) is given by the first moment
\begin{eqnarray}
A^{MBL}_{0}= \int_0^{+\infty} dw \Pi^{MBL}_{n}(w) w
\label{amplibeta0}
\end{eqnarray}

The power-law exponent in Eq. \ref{pktailpoisson}
is the same as the power-law tail expected in the Anderson Localized phase also governed
 by the Poisson statistics with $\beta=0$ \cite{edwards_thouless,mcmillan_anderson,kravtsov,montambaux,kramer,evangelou}.
However in localized phases, the question of the full distribution 
is much more tricky than the delocalized phase described by a simple expression (Eq. \ref{rmtbeta}) : 
in particular for the one-dimensional Anderson localization model,
the full distribution of Eq. \ref{pkfull} is not the Cauchy distribution 
(that would correspond to Eq. \ref{rmtbeta} for $\beta=0$), 
because the regular part $P^{reg}(k) $ 
describing the central part of the distribution 
has been found analytically \cite{fyodorov} and numerically 
\cite{kravtsov,montambaux,kramer,evangelou} to be log-normal.

In Ref. \cite{eisert}, the numerical results shown for the Many-Body-Localized phase
are actually rather similar : 
on their Fig. 1, the heavy tail seems in reasonable agreement with the tail
 exponent of Eq. \ref{pktailpoisson} if one takes into account finite-size effects,
 while on their Fig. 2, 
 the full distribution is not consistent with the Cauchy distribution,
and the central part of the distribution is well approximated by a log-normal distribution.

In summary, in the Many-Body Localized phase, it is essential to separate
the full distribution into the two contributions of Eq. \ref{pkfull} :

(i) the singular heavy-tail part $P^{sing}_{\beta}(k)$ for
 $\vert k \vert \to +\infty$ of Eq. \ref{pktailpoisson}
reflecting the absence of level repulsion $\beta=0$ of the Poisson statistics of the energy levels
is entirely due to the distribution of the term in the sum of Eq. \ref{Knper} 
corresponding to the smallest energy difference as explained in detail around Eq. \ref{knssmall}.

(ii) the regular part $P^{reg}(k) $ describing the central part of the distribution 
reflects the statistics of the whole sum over all eigenstates in Eq. \ref{Knper}
when this sum remains finite (i.e. not arbitrary large as in (i)).

\section{ Statistical properties of
 the fidelity susceptibility $\chi_n$ of eigenstates }

\label{sec_chi}

The fidelity susceptibility $\chi_n$ is dimensionless as a consequence of its definition (Eq. \ref{chifidel}),
so that no rescaling is needed in contrast to the level curvature (Eq. \ref{ksmall}).

\subsection{ Heavy-tails of the distribution of the fidelity susceptibility $\chi_n$ }

The mechanism discussed around Eq. \ref{knssmall}
 producing a very large curvature $k_n$ will similarly produce a very large fidelity susceptibility $\chi_n$ 
in Eq. \ref{chinper} :
 a very small eigenvalue spacing
$\vert E_{n}-E_{next} \vert =s \Delta_n  $ with $s \to 0$ in Eq. \ref{chinper},
yields a large fidelity susceptibility that can be approximated by
\begin{eqnarray}
\chi_n &&   \opsimeq_{ s \to 0} 
\frac{ \vert < \psi_{next}(0) \vert {\cal I}  \vert \psi_n(0) > \vert^2 }
{ \Delta_n^2 s^2  }
\label{chinssmall}
\end{eqnarray}
so the change of variable is slightly different from Eq. \ref{knssmall}.
As a consequence, the small-$s$ behavior of the probability distribution $R_{\beta}(s) \propto s^{\beta}$ (Eq. \ref{psbeta})
and the probability distribution $\Pi_{n}(w_n)$ of the variable $w_n$ introduced in Eq. \ref{wn}
yield by the change of variables of Eq. \ref{chinssmall}
\begin{eqnarray}
Q^{sing}_{\beta}(\chi)  && \opsimeq_{ \chi \to +\infty} \int_0^{+\infty} dw \Pi_{n}(w) \int_0^{+\infty} ds R_{\beta}(s)
\delta\left( \chi -\frac{ w } {  s^2  } \right)
\nonumber \\
&& \opsimeq_{ \chi \to +\infty} \int_0^{+\infty} dw \Pi_{n}(w) \int_0^{+\infty} ds R_{\beta}(s)
\frac{ \delta\left( s- \sqrt{ \frac{ w } { \chi  }  }\right) }{ \frac{ 2 w}{ s^3} }
\nonumber \\
&& \opsimeq_{ \chi \to +\infty} \int_0^{+\infty} dw \Pi_{n}(w)   \frac{ w^{\frac{1}{2} }} { 2  \chi^{\frac{3}{2}} }
  R_{\beta} \left( \sqrt{  \frac{ w } {  \chi  } } \right)
\nonumber \\
&&  \opsimeq_{ \chi \to +\infty} \int_0^{+\infty} dw \Pi_{n}(w)   \frac{ w^{\frac{1}{2} }} { 2  \chi^{\frac{3}{2}} }
  \left( \sqrt{  \frac{ w } {  \chi  } } \right)^{\beta}
\nonumber \\
&& \opsimeq_{ \chi \to +\infty}    \frac{ B_{\beta} } {   \chi^{\frac{3+\beta }{2}} }
\label{pchitail}
\end{eqnarray}
with the amplitude
\begin{eqnarray}
B_{\beta}=\frac{1}{2}  \int_0^{+\infty} dw \Pi_{n}(w)    w^{\frac{1+\beta}{2} }
\label{Bamplibeta}
\end{eqnarray}
involving the moment of order $\frac{1+\beta}{2}  $ of the random variable of Eq. \ref{wn}.

Again, it should be stressed that this analysis based only on the statistics on the nearest-energy-lvel term in the sum of Eq. \ref{chinper}
allows to predict the asymptotic power-law of Eq. \ref{pchitail} for $\chi \to +\infty $,
but a priori does not give any information on the shape of the distribution for finite values of $\chi$.
So in general, it is again convenient to decompose the full distribution $Q^{full}(\chi)  $
 into some unknown regular part $Q^{reg}(\chi) $ 
describing the central part of the distribution 
and into the singular power-law tail $Q^{sing}_{\beta}(\chi) $ of Eq. \ref{pchitail}
\begin{eqnarray}
Q^{full}(\chi) = Q^{reg}(\chi)+Q^{sing}_{\beta}(\chi)
\label{qchifull}
\end{eqnarray}

\subsection{ Many-Body-Delocalized ergodic phase $\beta=1$ } 

In the Many-Body-Delocalized ergodic phase governed by the GOE level statistics,
 the linear level repulsion exponent $\beta^{GOE}=1$ yields the following singular power-law tail of Eq. \ref{pchitail} 
\begin{eqnarray}
Q^{sing}_{ergodic(\beta=1)}(\chi)  \opsimeq_{\chi \to +\infty}  \frac{ B^{ergo}_{1} } {   \chi^2 }
\label{qktailgoe}
\end{eqnarray}
where the amplitude (Eq. \ref{Bamplibeta}) is given by the first moment
\begin{eqnarray}
B^{ergo}_{1}=\frac{1}{2}  \int_0^{+\infty} dw \Pi^{ergo}_{n}(w)    w
\label{Bamplibetagoe}
\end{eqnarray}

For Random matrices with $\beta=1,2$, exact results on the full probability distribution $Q^{full}(\chi) $ of the fidelity
can be found in \cite{fidelity_chaos}. As for the level curvature, one can expect that this RMT full distribution corresponding to $\beta=1$
will also describe the full distribution in the Many-Body-Delocalized ergodic phase.

\subsection{ Many-Body-Localized phase $\beta=0$  } 

In the Many-Body-Localized phase characterized by the Poisson level statistics, 
the absence of level repulsion $\beta=0$  yields the following singular power-law tail of Eq. \ref{pchitail} 
\begin{eqnarray}
Q^{sing}_{MBL(\beta=0)}(\chi)  \opsimeq_{\chi \to +\infty}    \frac{ B^{MBL}_{0} } {   \chi^{\frac{3}{2}} }
\label{qktailpoisson}
\end{eqnarray}
where the amplitude (Eq. \ref{Bamplibeta}) is given by the moment of index $1/2$
\begin{eqnarray}
B^{MBL}_{0}= \frac{1}{2}  \int_0^{+\infty} dw \Pi^{MBL}_{n}(w)    w^{\frac{1}{2} }
\label{Bamplibeta0}
\end{eqnarray}

As a final remark,
let us mention that the fidelity susceptibility $\chi_n$ 
of Eq. \ref{chinper} can be seen as the special case $q=1$ of the 
sums depending on the continuous parameter $q>0$ 
\begin{eqnarray}
\Sigma_q(n) \equiv  \sum_{k \ne n}  \left\vert\frac{  < \psi_k(0) \vert {\cal I}  \vert \psi_n (0)> } {E_n(0)-E_k(0)} \right\vert^{2q}
\label{sigmaq}
\end{eqnarray}
Similar sums have been studied in detail for the Anderson Localization power-law hopping model \cite{us_strongmultif}
and for some MBL quantum spin chain toy model \cite{c_mblper}.

\section{ Scaling with the system size $L$ of the amplitudes of the heavy tails }

\label{sec_ampli}

Besides the power-law exponents governing the heavy tails of the probability distributions of the level curvature (Eq. \ref{pktail})
and of the fidelity (Eq. \ref{pchitail}),
it is now important to discuss the dependence with respect to the system size $L$
of the amplitudes $A_{\beta}$  (Eq. \ref{amplibeta}) and $B_{\beta}$ (Eq. \ref{Bamplibeta}) of these heavy tails.
These amplitudes involves various moments of the variable $w_n$ introduced in Eq. \ref{wn}.
The statistics of matrix elements in Many-Body-Localization models
has been analyzed in various studies concerning the sensitivity to local perturbations 
 \cite{serbyn_criterion,c_mblpseudocriti,c_mblrgeigen,serbyn_dyson,c_dysonBM}.
In the present case, the variable $w_n$ introduced in Eq. \ref{wn}
 involving the matrix element
of the local operator $ {\cal I} $ between two consecutive eigenstates
has been studied in detail in Ref \cite{serbyn_criterion} in terms of the notation
\begin{eqnarray}
{\cal G} \equiv \ln \left( \frac{ \vert < \psi_{next}(0) \vert {\cal I}  \vert \psi_n(0) > \vert }{\Delta_n}
\right) =  \ln \sqrt{ w_n }
\label{calG}
\end{eqnarray}

Here it is important to emphasize that the the matrix element $\vert < \psi_{next}(0) \vert {\cal I}  \vert \psi_n(0) > \vert$ is between the level $n$
 and its nearest energy-level : 
these two levels have thus the same energy density $e=\frac{E_n}{L}$
and live in the same Hilbert space of effective dimension 
\begin{eqnarray}
{\cal N} \equiv {\cal N}_L(e) \simeq   \frac{1}{\Delta(E_n=Le) }   \opsimeq_{e \to 0} \frac{1}{ \sqrt { 2  \pi L \sigma^2 } }   e^{L \left( \ln 2 -  \frac{ e^2}{2  \sigma^2}  \right) }
\label{hilberte}
\end{eqnarray}
in terms of the level spacing discussed previously (Eqs \ref{largedev} and \ref{gaussdelta}).
So the question is how the moments of the variable $w_n$ of Eq. \ref{wn}.
\begin{eqnarray}
w_n && =  \frac{ \vert < \psi_{next}(0) \vert {\cal I}  \vert \psi_n(0) > \vert^2 }{\Delta_n^2} = {\cal N}^2 \ \ \vert < \psi_{next}(0) \vert {\cal I}  \vert \psi_n(0) > \vert^2
\label{wnbis}
\end{eqnarray}
scale with respect to the size ${\cal N} $ of the effective Hilbert space.

 \subsection{ Many-Body-Delocalized ergodic phase }

In the Many-Body-Delocalized ergodic phase, eigenvectors can be considered as random in
the effective Hilbert space of size ${\cal N}$ of Eq. \ref{hilberte},
so that the probability distribution of matrix elements of local operators 
can be explicitly computed (see for instance Eq A20 in the Appendix A of \cite{c_dysonBM}) :
for large ${\cal N}$, the probability distribution of $v=< \psi_{next}(0) \vert {\cal I}  \vert \psi_n(0) >$ of Eq. \ref{wnbis} becomes the Gaussian 
\begin{eqnarray}
P_{gauss}(v) \simeq \sqrt{ \frac{\cal N}{ 2 \pi}} e^{- \frac{{\cal N} v^2}{2  }}
\label{gauss}
\end{eqnarray}
For the variable $w={\cal N}^2 v^2 $ of Eq. \ref{wnbis}, this Gaussian form translates into the gamma-distribution of index $1/2$
\begin{eqnarray}
\Pi^{GOE}(w) \simeq \frac{1}{\sqrt{2 \pi {\cal N} w }} e^{- \frac{w}{2 {\cal N} }}
\label{pwgoe}
\end{eqnarray}
The moments of arbitrary index $q>0$ thus all diverge exponentially with respect to the system size $L$ 
\begin{eqnarray}
\int_0^{+\infty} dw w^q \Pi^{GOE}(w) && \simeq \frac{\Gamma \left( q+\frac{1}{2} \right) }{\sqrt{ \pi }} (2 {\cal N} )^q
\nonumber \\
&& \simeq \frac{\Gamma \left( q+\frac{1}{2} \right) }{\sqrt{ \pi }} \left( \sqrt{\frac{2}{   \pi L \sigma^2 } }   e^{L \left( \ln 2 -  \frac{ e^2}{2  \sigma^2}  \right) }\right)^q
\label{wqgoe}
\end{eqnarray}
in a very simple manner : the scaling of the moment $<w^q>$ of order $q$ is the same as the scaling of the $q$-power $(<w>)^q$ of the first moment $(<w>)$.

In particular, the amplitudes of Eq. \ref{amplibeta1}
and Eq \ref{Bamplibetagoe} can be obtained from the two special cases $q=2$ and $q=1$ respectively as
\begin{eqnarray}
A_{1}^{ergo}= \int_0^{+\infty} dw \Pi_{n}^{ergo}(w) w^2  \simeq \frac{\Gamma \left( q+\frac{1}{2} \right) }{\sqrt{ \pi }} \left( \sqrt{\frac{2}{   \pi L \sigma^2 } }   e^{L \left( \ln 2 -  \frac{ e^2}{2  \sigma^2}  \right) }\right)^2
\label{amplibeta1res}
\end{eqnarray}
and
\begin{eqnarray}
B^{ergo}_{1}=\frac{1}{2}  \int_0^{+\infty} dw \Pi^{ergo}_{n}(w)    w
\simeq \frac{\Gamma \left( q+\frac{1}{2} \right) }{ 2 \sqrt{ \pi }} \left( \sqrt{\frac{2}{   \pi L \sigma^2 } }   e^{L \left( \ln 2 -  \frac{ e^2}{2  \sigma^2}  \right) }\right)
\label{Bamplibetagoeres}
\end{eqnarray}
 From a physical point of view, the exponential divergence of these amplitudes with the system size $L$ in the ergodic phase reflects
 the chaotic nature of the spectrum which is extremely sensitive to any small local perturbation.

To make the link with Ref \cite{serbyn_criterion}, it is now interesting to write that the probability distribution $\Pi^{GOE}(w) $ of Eq. \ref{pwgoe}
translates for the variable ${\cal G}  = \ln \sqrt{ w_n } $ (Eq. \ref{calG}) into the Gumbel distribution
\begin{eqnarray}
{\cal P}^{GOE} ({\cal G} ) = \frac{2}{\sqrt \pi} e^{{\cal G}-{\cal G}_0 -e^{ 2({\cal G}-{\cal G}_0) }  }
\label{pcalGgoe}
\end{eqnarray}
where the only dependence on the size $L$ is in the characteristic scale
\begin{eqnarray}
{\cal G}_0 \equiv \frac{1}{2} \ln (2 {\cal N} ) = \frac{\left( \ln 2 -  \frac{ e^2}{2  \sigma^2}  \right)}{2} L
- \frac{1}{4} \ln \left( \frac{    \pi  \sigma^2 L } {2}   \right) 
\label{pcalGgoe0}
\end{eqnarray}
So as $L$ grows, the probability distribution of Eq. \ref{pcalGgoe} moves as a traveling-wave with a fixed shape without deformation,
in agreement with the numerical results displayed on Fig 2a of Ref. \cite{serbyn_criterion}
(For completeness, it would be interesting to check that the asymmetric shape of the traveling-wave on Fig 2a of Ref. \cite{serbyn_criterion} indeed corresponds to the Gumbel form of Eq. \ref{pcalGgoe}).

 \subsection{ Many-Body-Localized phase }

In the Many-Body-Localized phase, it has been found numerically that the probability distribution of the variable ${\cal G}$ defined in Eq. \ref{calG}
was approximately Gaussian \cite{serbyn_criterion}
\begin{eqnarray}
{\cal P}^{MBL}({\cal G}) \simeq \frac{1}{\sqrt{2 \pi {\cal V} } }  e^{- \frac{ ({\cal G}-{\cal G}_{av} )^2 }{2  {\cal V}}}
\label{gaussian}
\end{eqnarray}
around some average value ${\cal G}_{av} $ that decays linearly with the system size $L$ \cite{serbyn_criterion}
\begin{eqnarray}
{\cal G}_{av} \simeq - \kappa L
\label{gaussianav}
\end{eqnarray}
where the coefficient $\kappa>0$ parametrizes the localized phase (the critical point corresponding to $\kappa_c=0$).
From a physical point of view, the typical exponential decay of the variable $w_n$ with the system size $L$ (Eq. \ref{calG})
\begin{eqnarray}
w_n^{typ} = e^{ 2 {\cal G}_{av} } \simeq e^{- \kappa L}
\label{wtyp}
\end{eqnarray}
can be understood in terms of the 
 complete set of the Local Integrals of Motions (LIOMS) \cite{emergent_swingle,emergent_serbyn,emergent_huse,emergent_ent,
imbrie,serbyn_quench,emergent_vidal,emergent_ros,c_emergent,emergent_rademaker} : two nearby eigenstates in energy have typically 
different LIOMS everywhere in the sample, so that the off-diagonal matrix element of the local current operator involves some tunneling through the entire system
and is thus exponentially suppressed (see \cite{serbyn_criterion} for a much more detailed discussion).

The variance ${\cal V} $ in Eq. \ref{gaussian} has been found to grow with $L$ \cite{serbyn_criterion}, and it is natural to expect that it is linearly
\begin{eqnarray}
{\cal V} \simeq \sigma^2 L
\label{gaussianvar}
\end{eqnarray}

For the variable $w=e^{2 {\cal G}}$ (Eq. \ref{calG}), the Gaussian distribution of ${\cal G}$ (Eq \ref{gaussian}) translates into the log-normal distribution
\begin{eqnarray}
\Pi^{MBL}(w) \simeq \frac{1}{2 w \sqrt{2 \pi {\cal V} } }  e^{- \frac{ (\frac{\ln w}{2}-{\cal G}_{av} )^2 }{2  {\cal V}}}
\label{lognormal}
\end{eqnarray}
The moments of arbitrary index $q>0$ then reads
\begin{eqnarray}
\int_0^{+\infty} dw w^q \Pi^{MBL}(w) && = \int_{-\infty}^{+\infty} d {\cal G} e^{2q {\cal G }} {\cal P}({\cal G})
 =\int_{-\infty}^{+\infty} d {\cal G} e^{2q {\cal G }} \frac{1}{\sqrt{2 \pi {\cal V} } }  e^{- \frac{ ({\cal G}-{\cal G}_{av} )^2 }{2  {\cal V}}}
\nonumber \\
&& =e^{2 q {\cal G}_{av} + 2 q^2 {\cal V} }  = e^{ (-2 q\kappa + 2 q^2  \sigma^2) L } 
\label{wqmbl}
\end{eqnarray}
So here the situation is completely different from the case of Eq. \ref{wqgoe} : the scaling of the moment $<w^q>$ of order $q$ is 
not the same as the scaling of the $q$-power $(<w>)^q$ of the first moment $(<w>$, and this property is called multifractality.
In the present case, the multifractality is Gaussian in the index $q$ (Eq. \ref{wqmbl}) as a consequence of the log-normal distribution of Eq. \ref{lognormal}.

In particular, the amplitudes of Eq. \ref{amplibeta0} and Eq. \ref{Bamplibeta0}
can be obtained for the special cases $q=1$ and $q=\frac{1}{2} $ respectively as
\begin{eqnarray}
A^{MBL}_{0}= \int_0^{+\infty} dw \Pi^{MBL}_{n}(w) w =e^{2 {\cal G}_{av} + 2  {\cal V} } 
\simeq e^{ (- 2\kappa + 2  \sigma^2) L} 
\label{amplibeta0res}
\end{eqnarray}
and
\begin{eqnarray}
B^{MBL}_{0}= \frac{1}{2}  \int_0^{+\infty} dw \Pi^{MBL}_{n}(w)    w^{\frac{1}{2} } = \frac{1}{2}  e^{ {\cal G}_{av} + \frac{ {\cal V} }{2}  } 
\simeq  \frac{1}{2}  e^{ ( - \kappa  + \frac{ \sigma^2  }{2} ) L } 
\label{Bamplibeta0res}
\end{eqnarray}
These amplitudes  are exponentially decaying with the system size $L$ if $\kappa$ is sufficiently large (i.e. sufficiently deep in the Many-Body-Localized phase),
but can become exponentially growing with the system size $L$ if $\kappa$ is too small (i.e. sufficiently close to the Many-Body-Localization transition corresponding to $\kappa_c=0$).

\section{ Conclusion }

\label{sec_conclusion}

In summary, the sensitivity of the spectrum to twisted boundary conditions 
has been analyzed for one-dimensional Many-Body-Localization models, with the following conclusions :

(i)  the probability distribution $P_{\beta}(k)$ of the Thouless dimensionless curvature $k_n$ of the eigenvalues and the probability distribution $Q_{\beta}(\chi)$ of the fidelity susceptibility $\chi_n$ of the eigenstates both display heavy tails, and the corresponding power-law exponents involve the level repulsion exponent $\beta$ that takes the values $\beta^{GOE}=1$ in the ergodic phase and $\beta^{loc}=0$ in the localized phase. So this difference in the tail exponents can be used to locate the Many-Body-Localization transition, as found numerically in \cite{eisert}.

(ii) the amplitudes of these two probability distributions are given by moments of the off-diagonal matrix element of the local current operator between two nearby energy levels, whose probability distribution has been proposed as a criterion for the Many-Body-Localization transition \cite{serbyn_criterion}. These amplitudes display exponential 
behavior with respect to the system size $L$.

\end{document}